\documentclass[Journal]{IEEEtran}
\IEEEoverridecommandlockouts 
\usepackage{amsmath,amssymb,amsfonts}
\usepackage{dirtytalk}
\usepackage{graphicx}
\usepackage{cite}
\usepackage[keeplastbox]{flushend} 
\usepackage{tikz}
\usepackage{caption}
\usepackage{balance}
\usepackage{subcaption}
\usepackage{epstopdf} 
\usetikzlibrary{automata, positioning}
\hfuzz 
\vfuzz
\hbadness 
\vbadness
\newcommand{\RN}[1]{%
\textup{\uppercase\expandafter{\romannumeral#1}}
}
\begin{document}
\title{Optical Satellite Eavesdropping}
\author{Olfa Ben Yahia, \textit{Graduate Student Member},  Eylem Erdogan, \textit{Senior Member}, \textit{IEEE}, Gunes~Karabulut~Kurt, \textit{Senior Member}, \textit{IEEE}, 
Ibrahim Altunbas, \textit{Senior Member}, \textit{IEEE}, Halim Yanikomeroglu, \textit{Fellow}, \textit{IEEE}%
\thanks{O. Ben Yahia and I. Altunbas are with the Department of Electronics and Communication Engineering, Istanbul Technical University, Istanbul, Turkey, (e-mails: \{yahiao17, ibraltunbas\}@itu.edu.tr).}%
\thanks{E. Erdogan is with the Department of Electrical and Electronics Engineering, Istanbul Medeniyet University, Istanbul, Turkey, (e-mail: eylem.erdogan@medeniyet.edu.tr). }%
\thanks{G.~Karabulut~Kurt is with the Department of Electrical Engineering, Polytechnique Montréal, Montréal, QC, Canada (e-mail: gunes.kurt@polymtl.ca). }%
\thanks{H. Yanikomeroglu is with the Department of Systems and Computer Engineering, Carleton University, Ottawa, ON, Canada, (e-mail: halim@sce.carleton.ca).}}
\maketitle
\begin{abstract}
In recent years, satellite communication (SatCom) systems have been widely used for navigation, broadcasting application, disaster recovery, weather sensing, and even spying on the Earth. As the number of satellites is highly increasing and with the radical revolution in wireless technology, eavesdropping on SatCom will be possible in next-generation networks. In this context, we introduce the satellite eavesdropping approach, where an eavesdropping spacecraft can intercept optical communications established between a low Earth orbit satellite and a high altitude platform station (HAPS). Specifically, we propose two practical eavesdropping scenarios for satellite-to-HAPS (downlink) and HAPS-to-satellite (uplink) optical communications, where the attacker spacecraft can eavesdrop on the transmitted signal or the received signal. To quantify the secrecy performance of the scenarios, the average secrecy capacity and secrecy outage probability expressions are derived and validated with Monte Carlo simulations. Moreover, secrecy throughput of the proposed models is investigated. We observe that turbulence-induced fading significantly impacts the secrecy performance of free-space optical communication. 

\end{abstract}
\section{Introduction}

Over the past few years, an explosion in demand for high data rates has precipitated the integration of non-terrestrial networks with terrestrial infrastructure. In fact, three layers of heterogeneous networks are considered for state-of-the-art next-generation wireless communications. The well-known vertical heterogeneous network (VHetNets) architecture is composed of a space network, an aerial network, and a terrestrial network \cite{9380673}. In this architecture, low Earth orbit (LEO) satellites, which orbit at altitudes of less than 2,000 km, have attracted broad interests from researchers. LEO satellites are expected to be the key enabler of the space layer in future wireless networks due to their potential to provide real-time communication with enhanced data rates and high coverage.

Two interacting sub-layers are envisioned as part of the aerial network layer. As described by \cite{9356529}, one sub-layer would be composed of unmanned aerial vehicle (UAV) nodes, and the second sub-layer would include high altitude platform station (HAPS) systems. A HAPS is defined as an aircraft positioned in the stratosphere at an altitude between 17-22 km. These platforms may be airships, airplanes, or balloons \cite{d2016high}. HAPSs provide a number of advantages over terrestrial networks in terms of higher capacity, better propagation performance, more enhanced coverage, lower costs, and a better quality of service~\cite{arum2020review}.

To connect all the elements of VHetNets architecture and provide broad coverage with high data rates, free-space optical (FSO) communication is an important enabler. FSO communication provides an efficient way of transmitting data through free space with the use of laser beams. Benefits of FSO communication include high bandwidth, an unlicensed spectrum, low power requirements, and better security \cite{viswanath2015}. However, FSO links are affected by scintillation, pointing errors, and beam wander can be effective in uplink (UL) optical communication~\cite{9319151}. 

One major challenge in non-terrestrial networks is to guarantee a secure exchange of information. Typically, communication security is guaranteed through encryption. In recent research, however, a physical layer security (PLS) has been considered as a relevant solution complementary to cryptographic-based schemes to secure information against eavesdroppers relying on channel characteristics \cite{9238951}. Lately, PLS has been introduced for FSO communication. In FSO communication, the eavesdropper should be positioned very close to the communicating peers, so that it can capture the beam that is reflected by the aerosols or gases in the atmosphere. However, if the eavesdropper blocks the laser beam while intercepting the information, the receiver can notice the leakage of power and thus stop the communication for security issues. An alternative scenario can occur when misalignment arises between the communicating peers due to severe pointing errors or diffraction. In such case, the beam spreads out and the eavesdropper may receive the refracted beam \cite{8885999}.
Even though optical communication can be intercepted, only a few number of papers have considered PLS performance.
{Y. Ai \textit{et al.} \cite{9250517} discussed the PLS performance of three realistic scenarios by assuming different eavesdropper locations. It was shown that atmospheric condition imposes less impact on the secrecy performance when the eavesdropper is placed near the transmitter. In \cite{7038129}, the cases in which the eavesdropper is placed in proximity to the receiver or at proximity to the transmitter are investigated. It was observed that when the eavesdropper is close to the legitimate transmitter, it can easily intercept the communication by taking advantage of the larger attenuation affecting the signal when propagating through the FSO channel. In \cite{saber2017secure}, the authors derived the closed-form expressions of average secrecy capacity (ASC), secrecy outage probability (SOP), and strictly positive secrecy capacity (SPSC) over Málaga atmospheric turbulence channels in the presence of pointing errors for conventional optical communications. Furthermore, \cite{trinh2020secrecy} evaluated the impact of an eavesdropper's position on the secrecy performance of terrestrial FSO systems. In particular, the closed-form
expressions of the outage secrecy capacity and SPSC for PLS performance are derived.
The authors in \cite{monteiro2018maximum} studied the PLS performance of FSO communication in terms of a recent proposed metric named effective secrecy throughput. Furthermore, under generalized misalignments and atmospheric turbulence conditions, the authors in \cite{zhu2016average} studied the ASC for the log-normal fading model. Similar to RF communication, it was shown that channel correlation is crucial for secrecy capacity in all scenarios. The threat introduced by an eavesdropping spacecraft has not yet been investigated in the literature.
}

For decades, satellites have been valuable for politics and military purposes. Known as
\say{eyes in the sky,} reconnaissance satellites have been orbiting the Earth for more than 40 years \cite{norris2007spies}. Specifically, surveying the Earth from space is an effective strategy in wars and gathering information about other countries \cite{norris2007spies}. With the revolution in space technology, we highlight the idea of eavesdropping on satellites for next-generation networks. To the best of the authors' knowledge, this threat has not yet been studied in the literature.

{The current literature demonstrates a growing interest in security issues for FSO communication systems. However, these works are limited to optical terrestrial communications. Only \cite{yahia2021} investigated the optical eavesdropping in non-terrestrial networks by considering both HAPS and UAV eavesdropping. Motivated by the significant gap of PLS performance for optical communications in non-terrestrial networks, we make the following contributions:}
\begin{itemize}
    \item We introduce a new approach for optical satellite eavesdropping, where {a spacecraft is eavesdropping on an LEO satellite. }
    \item We investigate the secrecy performance of optical communication under two realistic scenarios. At first, we assume a downlink (DL) communication between an LEO satellite and a HAPS node. Secondly, we consider an UL communication from the HAPS to the satellite. In both cases, we assume that the external observer (eavesdropper {spacecraft}), which intends to intercept the communication, is located very close to the satellite, within the convergence area of its optical beam. 
    \item We study the impact of power leakage and adverse stratospheric conditions on the secrecy performance.
 \item We obtain novel closed-form expressions of the secrecy outage probability (SOP) and average secrecy capacity (ASC) for exponentiated Weibull (EW) fading while considering the amount of power captured by the eavesdropping {spacecraft}. {In addition, secrecy throughput (ST) analysis is provided.}
\end{itemize}

The remainder of the paper is organized as follows. Channels and system model are presented in Section $\RN{2}$. In Section $\RN{3}$, the expressions for secrecy performance are provided. Numerical results are outlined in Section $\RN{4}$ followed with related discussion. Finally, Section $\RN{5}$ concludes the paper.

\begin{figure}[!t]
  \centering
    \includegraphics[width=3in]{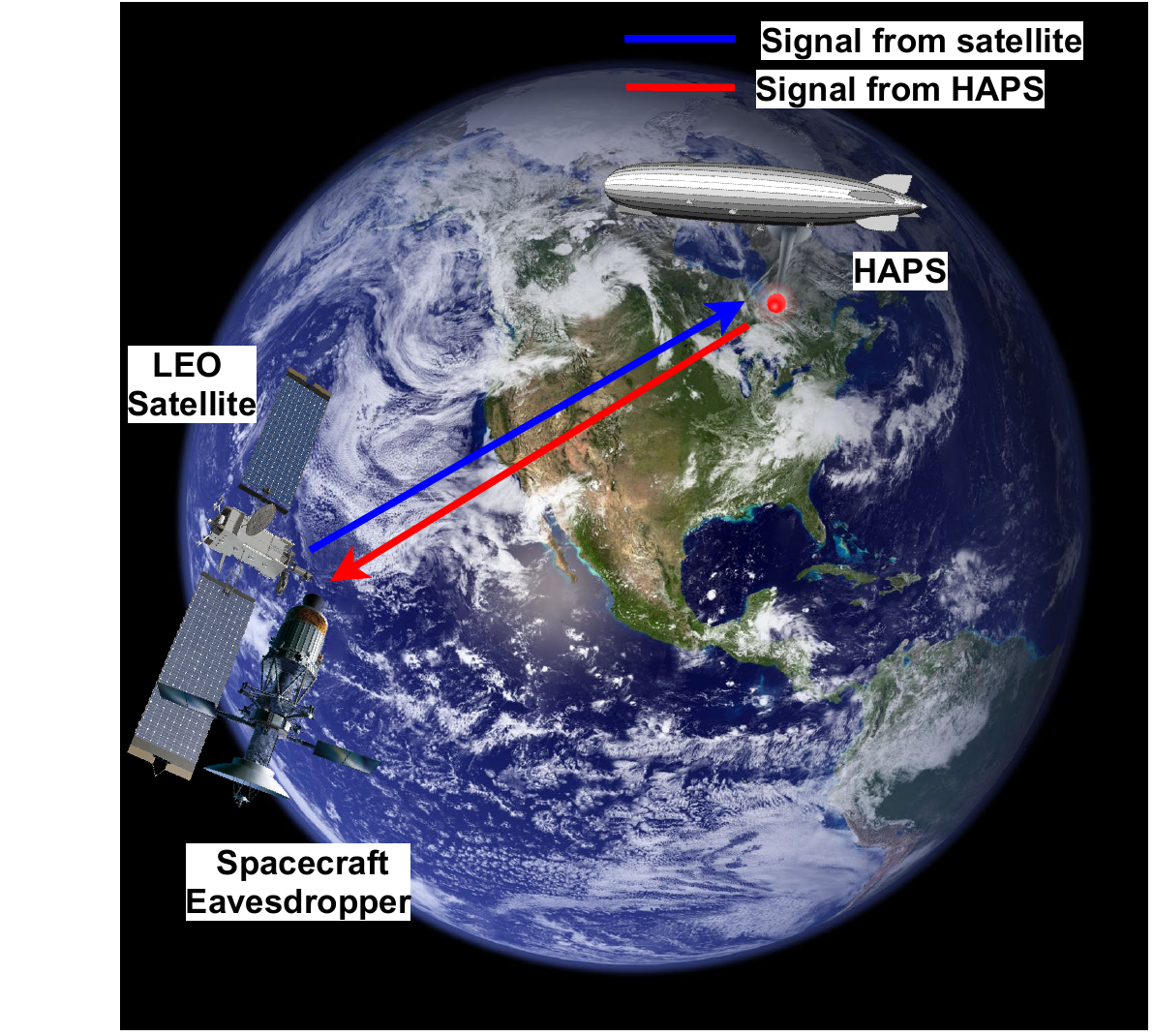}
    \caption{Illustration of the satellite eavesdropping.}
  \label{fig:systemModel}
\end{figure}
\section{Channels and System Model}
In this paper, we propose a novel scenario of eavesdropping attacks in space. As shown in Figure 1, we assume an LEO satellite $S$ that communicates with a HAPS $\mathcal{H}$ node in the presence of a sophisticated eavesdropping {spacecraft} $E$ located very close to $S$. We first consider $S$ to $\mathcal{H}$ DL communication, where $E$ is within the convergence area of the optical beam of $S$ so that it can eavesdrop on the transmitted beam. Secondly, we consider $\mathcal{H}$ to $S$ UL communication, where $E$ tries to capture the $\mathcal{H}$'s information. {Due to the HAPS node's quasi-stationary position, tracking and accuracy issues, and the Doppler shift can be tolerated for the communication between $S$ and $\mathcal{H}$.} In both scenarios, $E$ can capture only a small fraction $r_e$ of the transmitted signal, whereas the intended receiver collects more power resulting in a fraction $r_b$, where $r_e+r_b \leq 1$. Note that the parameters $r_b$ and $r_e$ depend on the aperture size of each device along with the beam divergence angle. {In the analysis, we assume that the locations of satellite, HAPS, and eavesdropper spacecraft can be used to extract the channel's physical model, which is representative of the channel state information \cite{andrews2005}.}

For $S$ to $\mathcal{H}$ communication scenario, the received signal at $\mathcal{H}$ can be given as $ y_\mathcal{H} = \sqrt{ r_b P_S} I_\mathcal{H} x_S + n_\mathcal{H}$, and for $\mathcal{H}$ to $S$ communication, the received signal at $S$ can be written similarly as $ y_S = \sqrt{ r_b P_\mathcal{H}} I_S  x_\mathcal{H} + n_S$. For both scenarios, the received signals at $E$ sent by node $j$, $j \in\left\lbrace S, \mathcal{H}\right\rbrace$ can be written as $ y_E = \sqrt{ r_e P_j} I_E  x_j + n_E$. $P_S$, $P_\mathcal{H}$ denote the transmit power of $S$ and $\mathcal{H}$ respectively, $I_\mathcal{H}$, $I_S $, $I_E$ indicate the received irradiance at $\mathcal{H}$, $S$, and $E$ respectively. $x_S$, $x_\mathcal{H}$ are the transmitted symbols with unit energy, and $n_\mathcal{H}$, $n_S$ indicate the additive white Gaussian noise (AWGN) with one-sided noise power spectral density $N_0$. Thus, the instantaneous signal-to-noise ratio (SNR) at $\mathcal{H}$ can be given as
\begin{align}
\label{SNR}
    \gamma_\mathcal{H}=\frac{r_b P_S} {N_0}{ I_\mathcal{H}^2 } =\overline{\gamma}_\mathcal{H} I_\mathcal{H}^2,
\end{align}
where $\overline{\gamma}_\mathcal{H}=\frac{r_b P_S }{N_0}$ defines the average SNR at $\mathcal{H}$ with $\mathbb{E}[I_\mathcal{H} ^2]=~1$, and the instantaneous SNR at $S$ and $E$ can be expressed similarly after changing the subscripts as $\gamma_\mathcal{S}=\frac{r_b P_\mathcal{H}} {N_0}{ I_S^2 } =\overline{\gamma}_S I_S^2$ and $\gamma_{E}=\frac{r_e P_j} {N_0}{ I_E^2 } =\overline{\gamma}_E I_E^2$ with $\mathbb{E}[I_E ^2] =\mathbb{E}[I_S ^2] = 1$. We assume that the turbulence-induced fading follows EW distribution. {EW fading has been shown to be the best fit for various aperture diameters for weak-to-strong turbulence levels, especially when aperture averaging is used to mitigate the impacts of turbulence and increase overall performance \cite{erdogan2020}.} Hence, the probability density function (PDF) of the irradiance $I$ for node $k$ where $k \in\left\lbrace S, \mathcal{H}, E\right\rbrace$ is expressed as in \cite [eqn. 11]{erdogan2020}.
Furthermore, the cumulative distribution function (CDF) of $\gamma_k$ can be expressed as \cite [eqn. 9]{8911487}
 \begin{align}
 \label{CDFEW}
 F_{\gamma_k}(\gamma)=\sum_{\rho=0}^{\infty}  \binom{\alpha_k}{\rho} 
   (-1)^{\rho} \exp\left[ -\rho \left( \frac{\gamma}{\eta_k^2 \overline{\gamma}_{k}} \right) ^{\frac{\beta_k}{2}}\right] ,
\end{align}
where $\alpha_{k}$, $\beta_{k}$, and $\eta_{k}$ indicate the shape parameters and the scale parameter, respectively, which depend on the scintillation index \cite{erdogan2020}. For the DL communication, the scintillation index $\sigma_{I_{DL}}^2$ at $\mathcal{H}$ can be written as
\begin{align}
\sigma_{I_{DL}}^2= \exp \left[ \frac{0.49 \sigma_{R}^2}{(1+1.11 \sigma_{R}^{\frac{12}{5}})^{\frac{7}{6}}} + \frac{0.51 \sigma_{R}^2}{(1+0.69 \sigma_{R}^{\frac{12}{5}})^{\frac{5}{6}}} \right] -1,
\label{EQN:9}
\end{align}
where $\sigma_{R}^2$ denotes the Rytov variance given in \cite[Sect. (12)]{andrews2005}, and it depends on the zenith angle $\xi_\mathcal{H}$ and the wind speed $w_\mathcal{H}$. In UL communication, beam wander induced-pointing error has to be taken into consideration \cite{yahia2021performance}. Therefore, the scintillation index $\sigma_{I_{UL}}^2$ can be written as
\begin{align}
\label{sigma}
   & \sigma_{I_{UL}}^2=5.95 (H_p-h_0)^2 \sec^2(\xi_p)\Bigg( \frac{2W_0}{r_0}\Bigg)^{\frac{5}{3}}  \Bigg( \frac{\alpha_{pe}}{W_p}\Bigg)^{2} \nonumber\\ 
    & + \exp \left[ \frac{0.49 \sigma_{Bu}^2}{(1+(1.11 +\Theta)\sigma_{Bu}^{\frac{12}{5}})^{\frac{7}{6}}} + \frac{0.51 \sigma_{Bu}^2}{(1+0.69 \sigma_{Bu}^{\frac{12}{5}})^{\frac{5}{6}}} \right]-1,
\end{align}
where $H_p$ is the height of $S$ and $E$, $h_0$ is the HAPS altitude, and $\xi_p$ is the zenith angle for UL communication.
$W_0$ is the beam radius at $\mathcal{H}$, the Fried’s parameter $r_0$ depends on the wind speed $w_p$, $\alpha_{pe}$ indicates the beam wander-induced pointing errors variance, $W_p$ is the beam size at $S$ and $E$, and $\sigma_{Bu}^2$ is the Rytov variance in UL communication \cite{yahia2021performance}.
\vspace{-0.2cm}
\section{Secrecy Analysis}
In this section, we analyze the secrecy performance of the proposed model. More specifically, the closed-form expressions of ASC, SOP, and ST are derived for both scenarios. According to the information-theoretic definition, the secrecy capacity $C_s$ is the maximum achievable secrecy rate that can be expressed as follows:
 \begin{align}
 C_s= \begin{cases} \log_2(1+\gamma_j)- \log_2(1+\gamma_E)~, \gamma_j >\gamma_E \\ \\ 
0~,   \text{otherwise,}
 \end{cases}
  \end{align}
where $\gamma_j$ denotes the instantaneous SNR of the main receiver.
\subsection{Average Secrecy Capacity}
In the context of PLS, ASC is an important metric for evaluating the secrecy performance of active eavesdropping. 
\subsubsection{Eavesdropping in Downlink Communication}
In $S$ to $\mathcal{H}$ DL communication, we assume that $E$ is positioned very close to $S$. Therefore, the turbulence-induced fading can be neglected to provide a realistic model \cite{7038129}. Thus, ASC for DL eavesdropping can be obtained by averaging the ${C}_s$ as
\begin{align}
  \label{ASC}
   \overline{C}_s&= \frac{1}{\ln(2)}\mathbb{E} \Big[ \ln (1+\gamma_\mathcal{H})-\ln (1+\gamma_E)\Big].
   \end{align}
   By using Jensen's inequality, for high SNR values, the ASC becomes
   \begin{align}
  \overline{C}_s &\cong \frac{1}{\ln(2)} \Big[\ln\Big(1+\mathbb{E}\left[ \gamma_\mathcal{H}\right] \Big)- \ln \Big(1+\mathbb{E}[{\gamma}_E]\Big)\Big]\nonumber\\ 
      &\cong\frac{1}{\ln(2)} \Big[\ln\Big(1+ \frac{r_b P_S} {N_0} \Big)- \ln \Big(1+\frac{r_e P_S} {N_0} \Big)\Big].
\end{align}
\subsubsection{Eavesdropping in Uplink Communication}
In UL communication, we take the turbulence-induced fading into consideration. Thereby, $\overline{C}_s$ can be expressed as \cite{9139494}
\begin{align}
   \overline{C}_s= \frac{1}{\ln(2)}\int_0^\infty \frac{F_{\gamma_E}(\gamma)}{1+\gamma} \Big[1- F_{\gamma_j}(\gamma) \Big] d \gamma .
   \label{ASC1}
\end{align}
By substituting $F_{\gamma_E}(\gamma)$ and $F_{\gamma_j}(\gamma)$ in (\ref{ASC1}), the ASC for UL communication becomes
\begin{align}
\label{EqCs}
 \overline{C}_s &= \sum_{\rho=0}^{\infty} \left( \begin{array}{c} \alpha_E \\
 \rho
   \end{array}  \right) 
   (-1)^{\rho }  \int_0^\infty \frac{1}{1+\gamma}  \exp\left[ -\rho \left( \frac{\gamma}{\eta_E^2 \overline{\gamma}_{E}} \right) ^{\frac{\beta_E}{2}}\right]  d\gamma\nonumber\\
   &- \sum_{\rho=0}^{\infty} \sum_{t=0}^{\infty}   (-1)^{\rho }  (-1)^{t} \left( \begin{array}{c} \alpha_E \\
 \rho
   \end{array}  \right) 
   \left( \begin{array}{c} \alpha_S \\
 t
   \end{array}  \right) \int_0^\infty \frac{1}{1+\gamma}
  \nonumber\\
   & \times   \exp\left[ -\Bigg(\rho \left( \frac{\gamma}{\eta_E^2 \overline{\gamma}_{E}} \right) ^{\frac{\beta_E}{2}}+ t \left( \frac{\gamma}{\eta_S^2 \overline{\gamma}_{S}} \right) ^{\frac{\beta_S}{2}}\Bigg)\right] d\gamma.
\end{align}
In $\mathcal{H}$ to $S$ communication, $E$ should be located very close to the main receiver's photo aperture to gather some of the reflected signals due to atmospheric pressure or misalignment caused by the pointing errors. Thus, we assume that $E$ encounters the same fading conditions as the intended receiver. Based on this assumption, we consider that $\beta_S=\beta_E=\beta$, $\alpha_S=\alpha_E=\alpha$, and $\eta_S=\eta_E=\eta$.
Thereafter, using the following transformations ${(1+x)}^a=\frac{1}{\Gamma(-a)}G_{1,1}^{1,1} \left(x  \middle \vert \begin{array}{c} a+1 \\ 0 \end{array} \right)$, where $G_{p,q}^{m,n} \Big( x\hspace{0.1cm} \Big| \begin{matrix} a_1,...,a_p\\  b_1,...,b_q \end{matrix} \Big)$ denotes the Meijer G-function, and $\exp(-bx)=G_{0,1}^{1,0}\left(bx  \middle \vert \begin{array}{c} - \\ 0 \end{array} \right)$  into (\ref{EqCs}), and by changing the variables as $X=\rho \left( \frac{\gamma}{\eta^2 \overline{\gamma}_{E}} \right) ^{\frac{\beta}{2}}$, and with the aid of
\cite[eqn. 07.34.21.0013.01]{Wolform}, the final expression of ASC can be formulated as in (\ref{ASCEX}) given at the top of the next page.
\begin{figure*}
\begin{align}
\label{ASCEX}
  &\overline{C}_s = \frac{1}{\ln2}\sum_{\rho=0}^{\infty} \binom{ \alpha}{\rho} 
   (-1)^{\rho } \frac{2}{\beta} \eta^2 \overline{\gamma}_{E} \left( \frac{1}{\rho} \right)^{\frac{2}{\beta}} \int_0^\infty X^{\frac{2}{\beta}-1} G_{1,1}^{1,1} \left(\eta^2 \overline{\gamma}_{E}\left( \frac{X}{\rho} \right)^{\frac{2}{\beta}} \middle \vert \begin{array}{c} 0 \\ 0 \end{array} \right)  \times  G_{0,1}^{1,0}\left(X  \middle \vert \begin{array}{c} - \\ 0 \end{array} \right) dX \nonumber \\
    &-\frac{1}{\ln2} \sum_{\rho=0}^{\infty} \sum_{t=0}^{\infty}   (-1)^{\rho +t }   \binom{ \alpha}{\rho} \binom{ \alpha}{t} 
 \frac{2}{\beta} \left(\frac{1}{\rho  A+ t B} \right) ^{\frac{2}{\beta}} 
   \int_0^\infty Y^{\frac{2}{\beta}-1} G_{1,1}^{1,1} \left(\left(\frac{Y}{\rho  A+ t B}\right)^{\frac{2}{\beta}} \middle \vert \begin{array}{c} 0 \\ 0 \end{array} \right)  \times G_{0,1}^{1,0}\left(Y  \middle \vert \begin{array}{c} - \\ 0 \end{array} \right) dY \nonumber \\
   &= \frac{1}{\ln2} \frac{2\times 2^{\frac{2}{\beta}- \frac{1}{2}}}{(2\pi)^{\beta-\frac{1}{2}}} \eta^2 \overline{\gamma}_{S}   \sum_{t=1}^{\infty} \binom{ \alpha}{t} 
   (-1)^{t+1 } \left( \frac{1}{t} \right)^{\frac{2}{\beta}} G_{\beta+2,\beta}^{\beta,\beta+2}\left( 4 \times \Big( \frac{\eta^2 \overline{\gamma}_{S}}{t^{\frac{2}{\beta}}} \Big)^\beta \middle \vert \begin{array}{c} \Delta(\beta,0), \frac{1-(2/\beta)}{2},\frac{2-(2/\beta)}{2} \\ \Delta(\beta,0) \end{array}\right) -\frac{1}{\ln2} \frac{2\times 2^{\frac{2}{\beta}- \frac{1}{2}}}{(2\pi)^{\beta-\frac{1}{2}}} \nonumber \\
   & \times \sum_{\rho=1}^{\infty} \sum_{t=1}^{\infty}     \binom{ \alpha}{\rho} \binom{ \alpha}{t} (-1)^{\rho +t+2 }    \left(\frac{1}{\rho  A+ t B} \right) ^{\frac{2}{\beta}} G_{\beta+2,\beta}^{\beta,\beta+2}\left( 4 \times \left(\frac{1}{\rho  A+ t B} \right) ^2 \middle \vert \begin{array}{c} \Delta(\beta,0), \frac{1-(2/\beta)}{2},\frac{2-(2/\beta)}{2} \\ \Delta(\beta,0) \end{array}\right).   
\end{align}
\hrulefill
\normalsize
\end{figure*}
\normalsize 
In (\ref{ASCEX}), the constants $A$ and $B$ show $A=\left(\frac{1}{\eta^2 \overline{\gamma}_{E}} \right) ^{\frac{\beta}{2}}$, $B=\left( \frac{1}{\eta^2 \overline{\gamma}_{S}} \right) ^{\frac{\beta}{2}}$, and $Y=\gamma^{\frac{\beta}{2}} \bigg(\rho  A+ t B \bigg)$.
\vspace{-0.3cm}
\subsection{Secrecy Outage Probability}
Another important metric in PLS is SOP, which is the best fit for the scenario of passive eavesdropping, where the source does not have any information about $E$. More specifically, SOP occurs when $C_s$ falls below a predefined secrecy rate $R_s$. Therefore SOP can be expressed as $P_\text{SO}= \Pr[C_s	\leq R_s ]$.
\subsubsection{Eavesdropping in Downlink Communication}
As mentioned above, in DL communication, $E$ is located very close to $S$ and the SOP expression can be written as
\begin{align}
\label{PSO}
   P_\text{SO} &= \Pr[\gamma_\mathcal{H}\leq 2^{R_s} +  2^{R_s}\overline{\gamma}_E -1 ]  \nonumber \\
    &=F_{\gamma_\mathcal{H}}(2^{R_s} +  2^{R_s}\overline{\gamma}_E -1 ).
        \end{align}
Thereby, by using (\ref{PSO}) $ P_\text{SO}$ can be obtained as
\small
        \begin{align}
        \label{SOP_DL}
      P_\text{SO}=\sum_{\rho=0}^{\infty} \binom{\alpha_\mathcal{H}}{\rho} 
   (-1)^{\rho} \exp \left[ -\rho \left( \frac{2^{R_s} +  2^{R_s} \overline{\gamma}_E -1 }{\eta_\mathcal{H}^2 \overline{\gamma}_{\mathcal{H}}} \right) ^{\frac{\beta_\mathcal{H}}{2}} \right].
    \end{align}
      \normalsize
\subsubsection{Eavesdropping in Uplink Communication}
For UL communication, since we take into account turbulence-induced fading, the expression SOP can be written as \cite{9473708}
\begin{align}
\label{SOPex}
  P_\text{SO} & =\int_{0}^{\infty}  F_{\gamma_j} \left( \gamma \gamma_{th} + \gamma_{th} - 1 \right) f_{\gamma_E} \left( \gamma \right)  d\gamma  \nonumber \\
   &\simeq \int_{0}^{\infty}  F_{\gamma_j} \left( \gamma\gamma_{th}\right) f_{\gamma_E} \left( \gamma \right)  d\gamma.
\end{align}
Thus, for UL eavesdropping, the SOP is expressed on the basis of (\ref{SOPex}), where the PDF of $\gamma_E$ can be derived from (\ref{CDFEW}) with respect to $\gamma$ as
\begin{align}
\label{PDFEW}
    f_{\gamma_E}(\gamma)&=\frac{\alpha_E \beta_E \gamma^{\frac{\beta_E}{2}-1}}{2(\eta_E^2\overline{\gamma}_E)^\frac{\beta_E}{2}}\sum_{q=0}^{\infty} \left( \begin{array}{c} \alpha_E - 1 \\
 q
   \end{array} \right) 
   (-1)^{q} \nonumber \\
   &\times \exp \Bigg[-(q+1) \Big(\frac{\gamma}{\eta_E^2\overline{\gamma}_E}\Big)^\frac{\beta_E}{2}  \Bigg].
\end{align}
Then, by substituting (\ref{CDFEW}) and (\ref{PDFEW}) into (\ref{SOPex}), we obtain the equation given on the top of the next page as (\ref{SOPeq}).
\begin{figure*}
\begin{align}
\label{SOPeq}
P_\text{SO}&= \int_{0}^{\infty} \frac{\alpha_E \beta_E \gamma^{\frac{\beta_E}{2}-1}}{2(\eta^2\overline{\gamma}_E)^\frac{\beta_E}{2}}
    \sum_{\rho=0}^{\infty}  \sum_{q=0}^{\infty} \binom{\alpha_S}{\rho}
  \binom{\alpha_E-1}{q} (-1)^{q+\rho} \exp\left[ -\rho \left( \frac{\gamma \gamma_{th}}{\eta_S^2 \overline{\gamma}_{S}} \right) ^{\frac{\beta_S}{2}}\right]    \exp \Bigg[-(q+1) \Big(\frac{\gamma}{\eta_E^2\overline{\gamma}_E}\Big)^\frac{\beta_E}{2}  \Bigg] d \gamma \nonumber\\
     &= \frac{\alpha \beta}{2(\eta^2\overline{\gamma}_E)^\frac{\beta}{2}}
    \sum_{\rho=0}^{\infty}  \sum_{q=0}^{\infty}\binom{\alpha}{\rho}\binom{\alpha - 1}{q}(-1)^{q+\rho} \int_{0}^{\infty} \gamma^{\frac{\beta}{2}-1}\exp\left[ -\Bigg(\rho \left( \frac{\gamma_{th}}{\eta^2 \overline{\gamma}_{S}} \right) ^{\frac{\beta}{2}}+(q+1)\left( \frac{1}{\eta^2 \overline{\gamma}_{E}} \right) ^{\frac{\beta}{2}} \Bigg) \gamma^{\frac{\beta}{2}}\right] d\gamma.
\end{align}
\hrulefill
\end{figure*}
Finally, after some mathematical derivations and by using \cite[3.478.1]{2014table}, the final expression of SOP can be written as 
\begin{align}
\label{SOP_UL}
P_\text{SO}&=\frac{\alpha }{(\eta^2\overline{\gamma}_E)^\frac{\beta}{2}}
    \sum_{\rho=0}^{\infty} \binom{\alpha}{\rho}
   (-1)^{\rho} \sum_{q=0}^{\infty} \binom{\alpha - 1}{q}
   (-1)^{q} \nonumber \\
   & \times \Bigg( \rho \left( \frac{\gamma_{th}}{\eta^2 \overline{\gamma}_{S}} \right)+   (q+1)\left( \frac{1}{\eta^2 \overline{\gamma}_{E}} \right)   \Bigg)^{-\frac{\beta}{2}}.
\end{align}
{\subsection{Secrecy Throughput}
Another metric to evaluate the secrecy performance of the proposed system is ST. This metric is exploited to characterize the overall efficiency of achieving reliable and secure transmission \cite{9031693}.
Mathematically, it can be written as
\begin{align}
\label{ST}
    ST=R_s (1-P_\text{SO}).
\end{align}
The expressions of ST for uplink and downlink communication can be easily obtained by substituting (\ref{SOP_DL}) and (\ref{SOP_UL}) in (\ref{ST}).
}
\section{Numerical Results and Discussion}
In this section, we evaluate the secrecy performance of satellite eavesdropping for the proposed models. For both scenarios, we consider the same parameters. The LEO satellite orbits at an altitude of 500 km, while the HAPS is located at 18 km. The zenith angles are set to $\xi_S=\xi_\mathcal{H}=70^\circ$, the wind speed is given as $w_S=w_\mathcal{H}=65$ m/s as we consider non-static stratospheric winds, and the secrecy rate is set as $R_s=0.01$ bit/s/Hz for SOP simulations.

\begin{figure}[!t]
  \centering
    \includegraphics[width=3in]{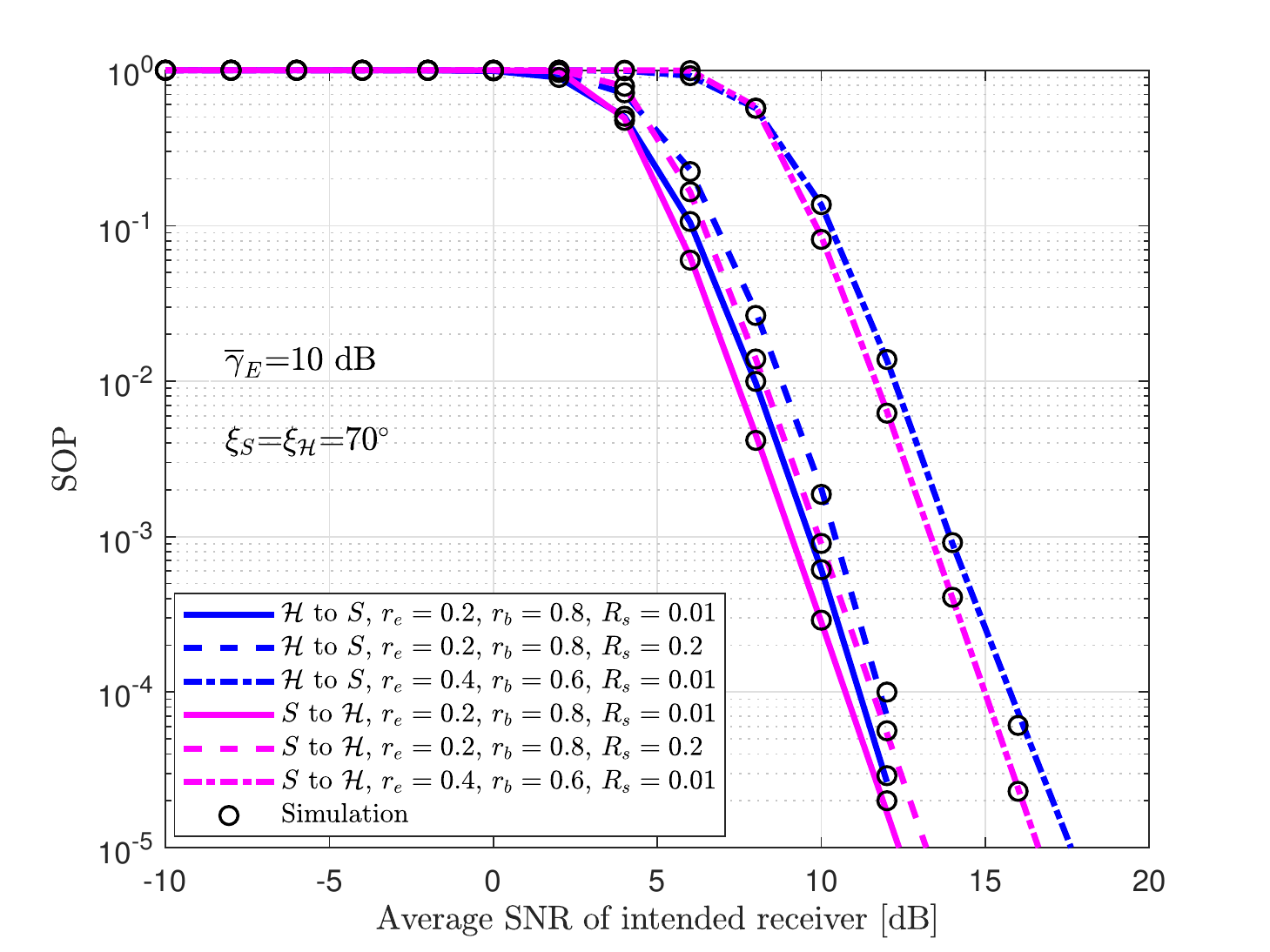}
    \caption{SOP performance of the proposed scenarios for different fractions of received power and different secrecy rate when $\overline{\gamma}_E=$10 dB.}
  \label{fig:SOP}
\end{figure}

\begin{figure}[!t]
\label{fi1}
  \centering
    \includegraphics[width=3in]{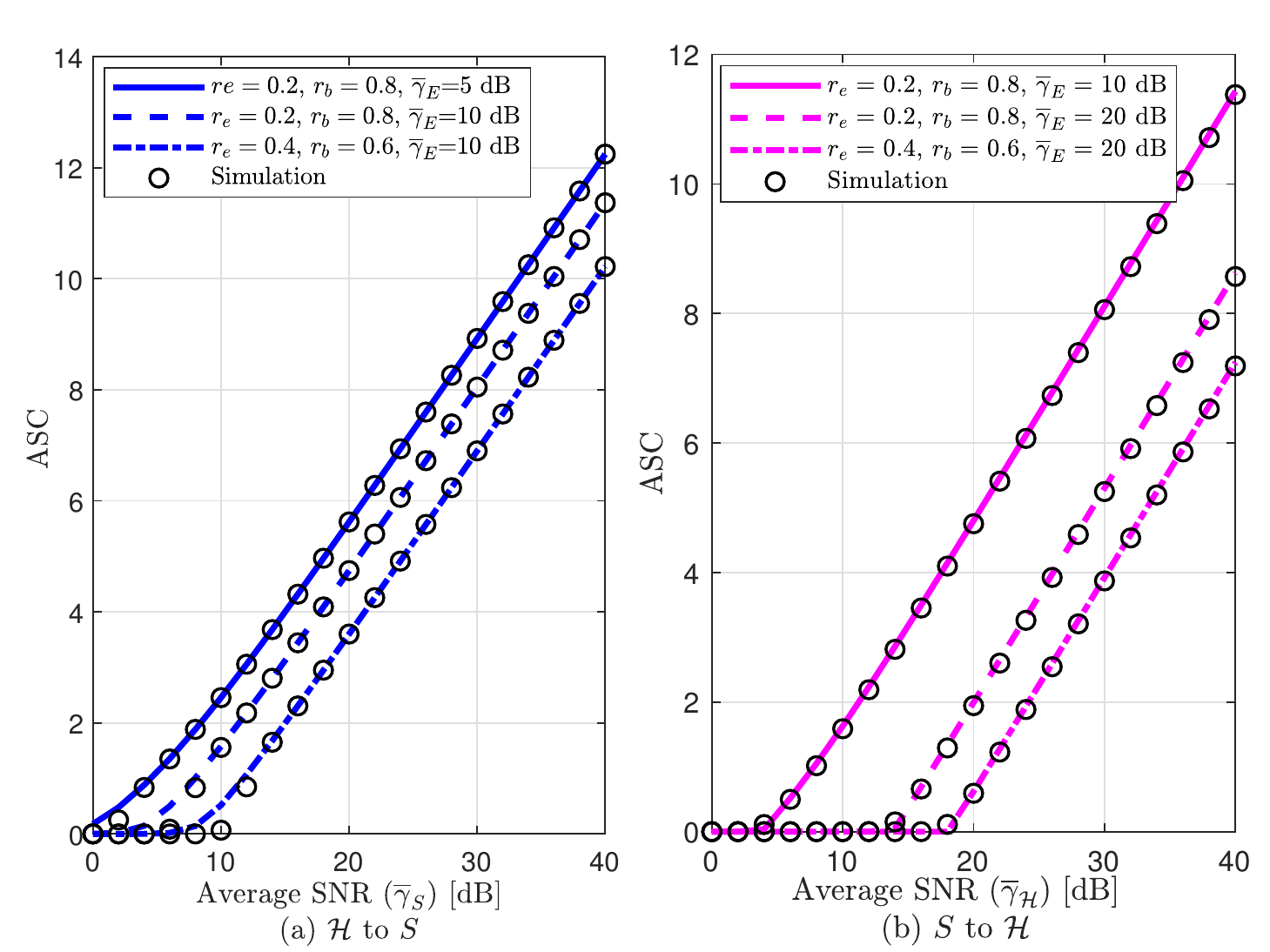}
    \caption{ASC performance of both scenarios for different $\overline{\gamma}_E$ and different fractions of received power.}
  \label{fig:ASC}
\end{figure}
In Figure 2, the SOP performance for both scenarios is plotted as a function of the SNR of the legitimate receiver. The eavesdropping average SNR is set to $\overline{\gamma}_E=10$ dB. As we can see, the secrecy performance of the $S$ to $\mathcal{H}$ communication is slightly better than that of $\mathcal{H}$ to $S$ communication.
This can be explained by the fact that when the illegitimate receiver approaches the receiver, it can collect more information as more beams can be reflected due to the greater distance. Moreover, for both scenarios, we can observe that the performance decreases as the fraction of the leaked power $r_e$ received by the eavesdropper increases. {Furthermore, as shown in the figure, increasing the secrecy rate $R_s$ deteriorates the SOP performance.} In addition, we can see the agreement of the theoretical expressions with the MC simulations.

In Figure 3, we see the ASC performance for both scenarios. As the figure shows, by increasing the average SNR of the eavesdropper $\overline{\gamma}_E$ or decreasing the amount of power captured by the legitimate receiver, the overall secrecy performance degrades. Furthermore, for the $S$ to $\mathcal{H}$ communication, we assume that the eavesdropper has a stronger SNR, as it is located very close to the LEO satellite and does not experience turbulence-induced fading effects. Additionally, it is clear from the figure that the analytical expressions match exactly with the MC simulations. Finally, we can observe that perfect secrecy is achieved when the average SNR of the intended receiver is higher than the SNR of the eavesdropper.

Figure 4 shows the impact of the zenith angle on the proposed satellite eavesdropping scenarios. We set $\overline{\gamma}_E=4$ dB and the fraction of power received by $E$ as $r_e=0.1$. As we can see, increasing the zenith angle decreases the amount of information that is captured by the legitimate receiver and leaked to the eavesdropper. Furthermore, the fluctuation in the signal is significant for a higher zenith angle. Thus, we conclude that there is a huge loss of performance. Moreover, as seen in the figure, for lower zenith angle, the SOP performance is almost the same for both scenarios, and the simulation results reveal that the scintillation indexes are equals. However, when increasing the zenith angle to $80^\circ$, we can see $2$ dB difference between the two curves. Also, the impact of beam wander increases with increasing zenith angle.
\begin{figure}[!t]
  \centering
    \includegraphics[width=3in]{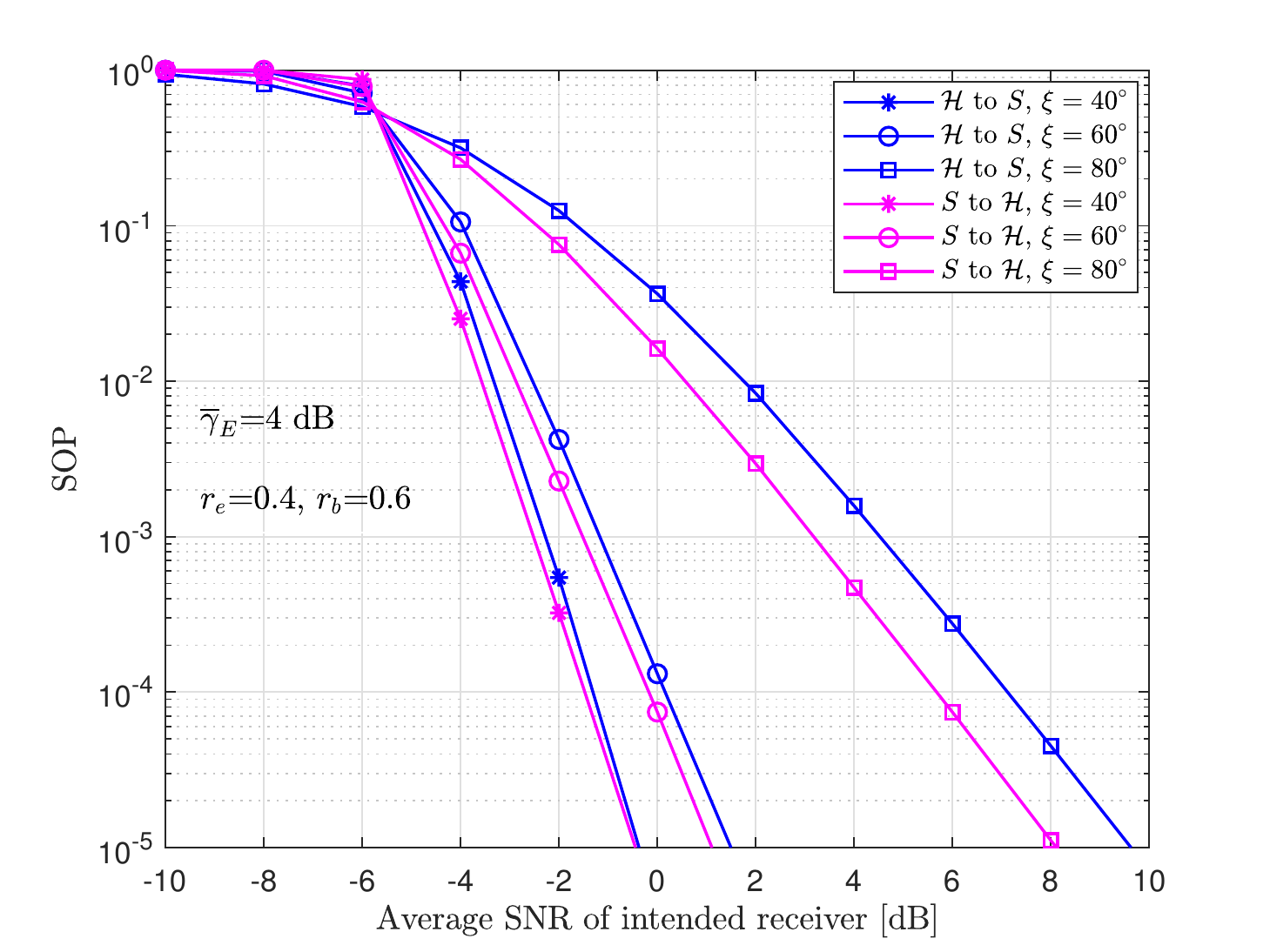}
    \caption{SOP performance of the proposed models under different zenith angles, $\overline{\gamma}_E=$4 dB.}
  \label{fig:SOP2}
\end{figure}

\begin{figure}[!t]
  \centering
    \includegraphics[width=3in]{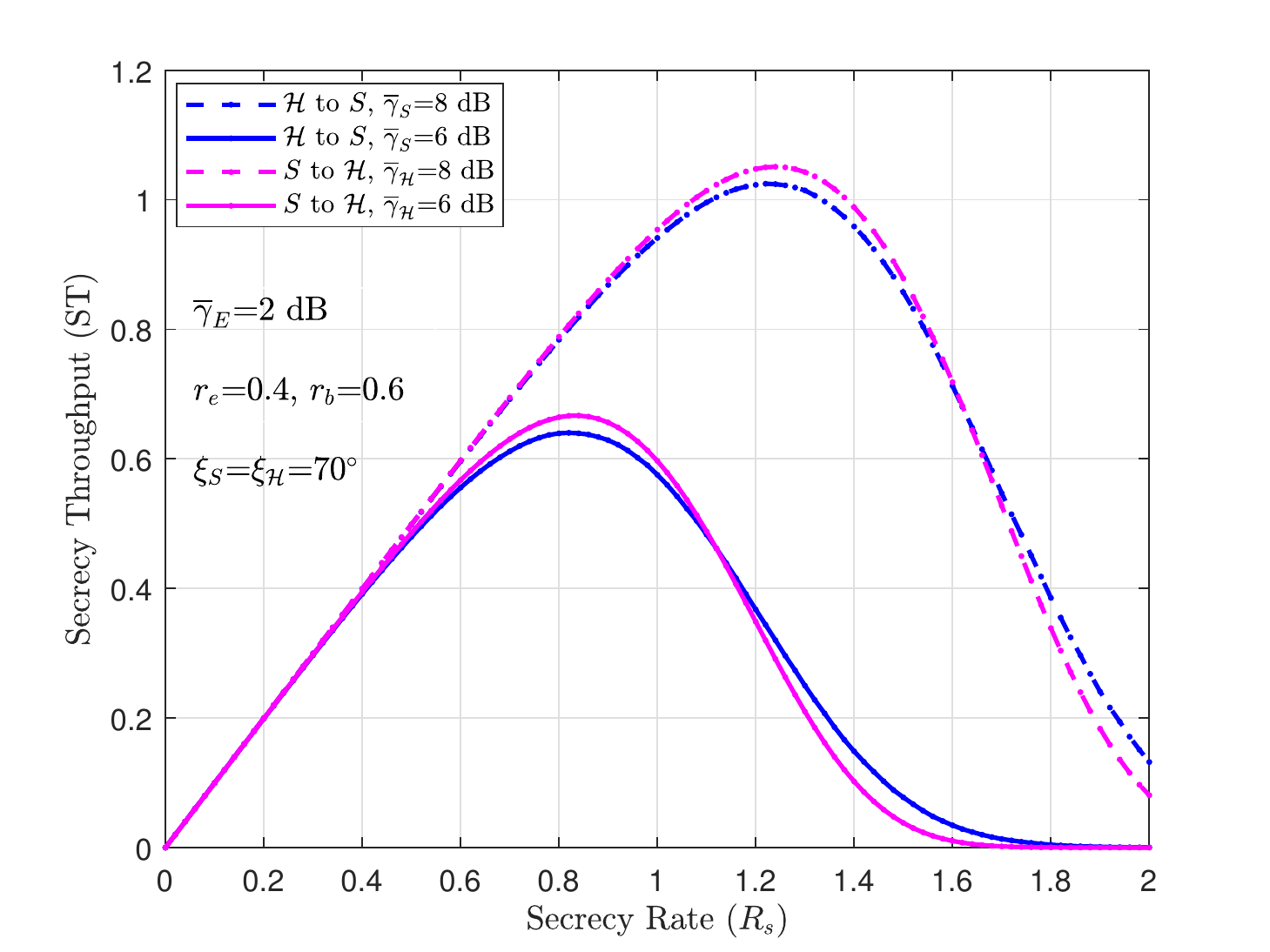}
    \caption{ST performance of the proposed models vs. $R_s$.}
  \label{fig:ST}
\end{figure}

{Figure \ref{fig:ST} evaluates the ST versus the target secrecy rate $R_s$ for the proposed models for different average SNR levels of the intended receiver. In all cases, the average SNR of the spacecraft is set to $2$ dB. As can be observed, the ST increases as $R_s$ increases to a certain value and then decreases. This is due to the dependency of ST on $R_s$. The ST increases when the $R_s$ is relatively low. However, when $R_s$ exceeds a particular threshold value, the system
cannot afford reliability and security. Furthermore, it is observed that the $S$ to $\mathcal{H}$ downlink communication performs slightly better than $\mathcal{H}$ to $S$ uplink scenario at lower $R_s$. However, after a particular value of $R_s$, the uplink scenario outperforms the downlink communication. Finally, as expected, increasing the average SNR of the intended receiver improves the ST performance.
}

Our main observations from the numerical results can be summarized as follows:
\begin{itemize}
    \item Increasing the zenith angle between the HAPS node and the satellite decreases the overall performance.
    \item In the UL scenario, the eavesdropper {spacecraft} can collect more information as more beams can be reflected due to the greater distance.
    \item The fluctuations in the signal caused by atmospheric conditions have a direct impact on the secrecy performance.
    \item The SNR of the eavesdropper and the amount of power leaked to the eavesdropper are critical parameters to provide secure communication.
    \item {From ST point view, it is observed that
the system reliability and secrecy are compromised after a certain value of secrecy rate $R_s$.}
\end{itemize}
{These observations can be used to provide practical insights about the
system architecture, so that a system designer can get a quick idea about the overall system
performance, especially, as we are considering random channel characteristics.}
\section{Conclusion and Future Works}
This paper has introduced the satellite eavesdropping approach, where {a spacecraft} eavesdrops on an LEO satellite. Specifically, we assumed an LEO satellite communicating with a HAPS node in the presence of {an attacker spacecraft} located very close to the LEO satellite. The unified expressions of SOP and ASC were derived in closed-form assuming EW fading. The expressions obtained were validated with MC simulations. In addition, the simulation results showed that satellite-to-HAPS downlink communication is more secure. It was also shown that increasing the leaked beam collected by the eavesdropping {spacecraft} deteriorated the secrecy performance. We observed that turbulence-induced fading highly affected the secrecy performance.

{The future work directions can be summarized as follows: PLS techniques against wireless eavesdroppers as wiretap coding will be studied while exploiting the channel characteristics along with the transceiver architecture in order to favor the data transmission between the legitimate users. Finally, optimization of the ST by adjusting the target secrecy rate and the power allocation ratio will be carried out.}

\balance
\bibliographystyle{IEEEtran}
\bibliography{refe}
\end{document}